\documentclass[prl,aps,twocolumn,showpacs]{revtex4}
\usepackage{graphicx,latexsym}
\usepackage{dcolumn}
\usepackage{amsmath,amssymb,epsf,bm}
\begin{document}
\title{Equilibrium Circular Photogalvanic Effect in a hybrid
superconductor-semiconductor system}
\author{A.~G. Mal'shukov}
\affiliation{Institute of Spectroscopy, Russian Academy of
Sciences, 142190, Troitsk, Moscow oblast, Russia}
\begin{abstract}
A dc electric current can be induced in a hybrid
semiconductor-superconductor system under illumination it by a
circularly polarized light with the frequency below the energy of
semiconductor interband transitions. In conditions when the light
beam is unable to create real electron-hole excitations, this
phenomenon is reminiscent of the Meissner effect in the static
magnetic field. Such an effect can be employed in systems combining
cavity photons and superconducting quantum circuits.
\end{abstract}
\maketitle

In noncentrosymmetric semiconductors a circularly polarized light
can generate the dc electric current  \cite{pherev}. This, so
called, circular photogalvanic effect (CPE) is determined by the
second-order response to the electric field of the incident light.
It involves spin polarization of electrons excited from valence
bands split by the spin-orbit interaction (SOI) into bands with the
total angular moments 3/2 and 1/2. The polarized carriers, in their
turn, give rise to the electric current, due to SOI associated with
the lack of inversion symmetry. This phenomenon attracted much
attention recently \cite{pheNature} in connection with perspective
spintronic applications.

As expected, in the case of a dissipative transport in normal
electron systems at stationary conditions, CPE takes place only if
light illumination creates a thermodynamically nonequilibrium
distribution of carriers in a semiconductor \cite{Belinicher}. On
this reason, one can not expect this effect to occur in bulk
semiconductors, if the frequency of the incident light is less than
the energy gap between the valence and conduction bands and when
indirect phonon/impurity assisted optical transitions are weak. On
the other hand, in superconductors the flow of electrons might be
created without exciting the system from its thermal equilibrium.
Such a phenomenon becomes possible due to a macroscopic coherency of
the condensate wave function, so that spacial variations of its
phase determine the supercurrent. The well known example is the
Meissner effect, where the static magnetic field can not produce
electron-hole excitations. It, however, induces a superconducting
electric current.  The goal of this Letter is to show that CPE can
be observed in a hybrid superconductor-semiconductor system even if
the incident light does not drive it from the thermal equilibrium.
The electric current is induced due to changes in the spectrum and
wave function of the many-electron system, rather than from
deviation of the electron's distribution function from its thermal
equilibrium. Apart from the fundamental interest, this phenomenon
has a practical value when it is employed in nanoscale
optoelectronic devices, because it allows to reduce dissipation
losses compare to CPE in normal systems.  It can also be an
efficient tool to combine microwave electrodynamics in
superconducting quantum circuits \cite{circuits} with cavity optical
fields coupled to atomic ensembles \cite{cQED}, like the recently
observed \cite{Schuster} interaction of spin ensembles to
superconductor resonators. This suggests evident connections to many
topics of current interest, ranging from atomic and polaritonic Bose
condensates to spintronics and quantum information processing.

The equilibrium CPE takes place in a hybrid
semiconductor-superconductor system due to the proximity effect
which induces Cooper pair correlations in the semiconductor. Various
hybrid systems displaying the strong proximity effect has been
recently fabricated \cite{InAs}. In order to demonstrate main
features of CPE and make our quantitative analysis more transparent,
a simple model will be considered here, where a noncentrosymmetric
semiconductor film is in a planar contact with superconductor having
the singlet order parameter. The film contains an n-doped quantum
well (QW) close to the metal surface (Fig. 1). Note, that such a
sandwich system is now being considered as a key element of a
topological superconductor \cite{Sarma}.

\begin{figure}[bp]
\includegraphics[width=6cm]{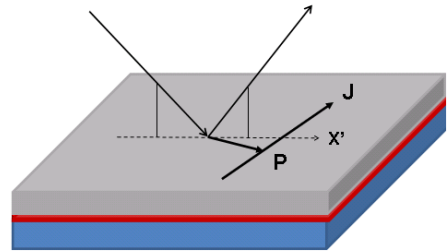}
\caption{(Colour online) A sketch of the system. An optically
transparent semiconductor film (top) is in a contact with a
superconductor (bottom). A thin layer between them depicts a doped
quantum well. An incident electromagnetic wave induces in this well
an electric current $\mathbf{J}$ perpendicular to the vector
$\mathbf{P} \sim \mathbf{E}\times \mathbf{E}^*$, where $\mathbf{E}$
is the electromagnetic field. The light beam can be also incident
from the metal side, if the metal film is thinner than the
skin-layer.} \label{fig1}
\end{figure}

Let us assume that the circularly polarized light with the frequency
$\omega_i$ is incident onto the semiconductor surface. The vector
potential of the total electromagnetic field in the QW region is
represented by its Fourier components $\mathbf{A}$ and
$\mathbf{A}^*$, corresponding to frequencies $\omega=\omega_i$ and
$\omega=-\omega_i$, respectively. It will be assumed that the
electromagnetic field causes only direct transitions between valence
and conduction bands. Hence, phonon and impurity assisted indirect
transitions, as well as a very small momentum transfer by photons,
will be ignored. The corresponding matrix elements are given by
$M_{\alpha v}(\omega_i)=(e/mc)\langle\mathbf{k},\alpha|
\mathbf{p}\cdot\mathbf{A}|\mathbf{k} , v \rangle$ and $M_{\alpha
v}(-\omega_i)=(e/mc)\langle\mathbf{k},\alpha|
\mathbf{p}\cdot\mathbf{A^*}|\mathbf{k} , v \rangle$ , where
$\mathbf{k}$ denotes the two-dimensional wave vector of QW electrons
and $\mathbf{p}$ is their momentum operator, $\alpha$ is the
conduction electron spin variable and $v$ labels the valence bands.
In 3D superconductors the latter are  the heavy-hole and light-hole
bands with the total angular moment $l=3/2$ and its projections $\pm
3/2$ and  $\pm 1/2$, respectively, plus the split-off band with
$l=1/2, l_z=\pm 1/2$. In QW the label $v$ runs also through valence
subband indexes, while only the lowest conduction subband is assumed
to be occupied. In noncentrosymmetric semiconductors SOI splits the
energies of bands with opposite angular moments. For CPE this
splitting is crucial. In the case of a n-doped QW the conduction
band splitting appears to be most important, although at strong
resonance conditions the hole splitting can be equally important.
Therefore, the latter will be ignored under the assumption that it
is much less than the resonance detuning. At the same time, the
conduction band splitting is determined by the SOI Hamiltonian
$H_{so}=\bm{\sigma}\cdot \mathbf{h}_{\mathbf{k}}$, where
$\bm{\sigma}=(\sigma_x, \sigma_y,\sigma_z)$ is the vector of Pauli
matrices and $ \mathbf{h}_{\mathbf{k}}=- \mathbf{h}_{-\mathbf{k}}$
is the spin-orbit field. The latter will be assumed  to have only
$x$ and $y$ components. This takes place for the Rashba field, as
well as for the Dresselhaus field in [001] oriented QW
\cite{Winkler}.

The semiconductor film contacts 3D metal through a high enough
tunneling barrier, so that the broadening of QW states due to
resonance with 3D continuum is much smaller than the metal
superconducting gap. Therefore, in the leading approximation, the
electric current in QW can be calculated by ignoring a leaking of
electrons into the metal. The system is assumed to be clean enough,
with the mean electron scattering rate in QW to be much smaller than
the proximity induced energy gap and $h_{\mathbf{k}}$. Hence, the
Green's functions are diagonal with respect to electron wave
vectors. Considering a thermodynamic equilibrium with the
temperature $T$, the stationary electric current density can be
expressed in terms of the equilibrium Keldysh function, as
\begin{eqnarray}\label{J}
\mathbf{J}&=&-\frac{ie}{4}\sum_{\mathbf{k},n}\int \frac{d\omega}{2\pi}Tr[\mathbf{j}_{n\mathbf{k}}(G^r_{n\mathbf{k}}(\omega)-G^a_{n\mathbf{k}}(\omega))] \times  \nonumber \\ && \tanh\frac{\omega}{2k_BT} \,,
\end{eqnarray}
where $G^{r(a)}_{n\mathbf{k}}(\omega)$ are the retarded and advanced
Green's functions. The index $n$ labels the energy bands including
conduction $n=c$ and valence $n=v$ bands. Although the valence bands
are far from the Fermi level, their  inclusion into the current
expression is necessary, because they contribute to important
spectral corrections associated with the coupling of electrons to
the electromagnetic field. The current operator of conduction
electrons is given by
\begin{equation}\label{j}
\mathbf{j}_{\mathbf{k}}= \frac{\partial \epsilon_{c\mathbf{k}}}{\partial \mathbf{k}} + \tau_3  \frac{\partial (\bm{\sigma}\cdot \mathbf{h}_{\mathbf{k}})}{\partial \mathbf{k}}\,,
\end{equation}
while the valence band currents are given by corresponding band
velocities $\partial\epsilon_{v,\mathbf{k}}/\partial \mathbf{k}$,
where $\epsilon_{n,\mathbf{k}}$ are the band energies measured with
respect to the chemical potential. The above expressions are written
in the Gor'kov- Nambu representation, where the new electron destruction
operators   $c_{\mathbf{k},n,\sigma, 1}=c_{\mathbf{k},n,\sigma}$ and
$c_{\mathbf{k},n,\sigma ,2}=c^{\dag}_{-\mathbf{k},\bar{n},
\bar{\sigma}}$ are introduced, with Pauli matrices $\tau_1, \tau_2,
\tau_3 $ acting in the space of indices 1 and 2. As mentioned in
Ref.\cite{Bergeret},  this basis is more convenient for
studying the spin-dependent transport in superconducting systems.

The next step is to specify important contributions to the electron
self-energy due to the superconducting proximity effect and
interaction with the electromagnetic field. The  proximity effect is
represented by a nondiagonal in the Nambu space contribution to the
self-energy of QW electrons. In the case when parallel to the
interface wave vectors of  tunneling particles are conserved, this
self-energy has the form
\begin{equation}\label{Sigma}
\Sigma^{r(a)}_t(\omega,\mathbf{k})= \sum_{k_z}|T_{\mathbf{k}, k_z}|^2G^{r(a)}_{\mathbf{k}, k_z}(\omega)\,,
\end{equation}
where $T_{\mathbf{k}, k_z}$  is the tunneling matrix element, with
$k_z$ denoting the normal component of the wave vector in the metal
side of the interface. Assuming the real order parameter $\Delta_0$,
the Green's function in (\ref{Sigma}) can be written  as
$G^{r(a)}_{\mathbf{k}, k_z}=(\omega-\tau_3\epsilon_k-\sigma_z
\tau_1\Delta_0\pm i\delta)^{-1}$.  For a wide-band metal, where
$T_{\mathbf{k}, k_z}$ slowly varies near the Fermi level, the
self-energy is expressed as
\begin{equation}\label{Sigma2}
\Sigma^{r(a)}_t(\omega,\mathbf{k})=-\Gamma_{\mathbf{k}}(0)\frac{\omega+\tau_1\sigma_z \Delta_0}{\sqrt{\Delta_0^2-(\omega \pm i\delta)^2}}\,,
\end{equation}
where $\Gamma_{\mathbf{k}}(\epsilon)= \sum_{k_z}|T_{\mathbf{k},
k_z}|^2\delta (\epsilon-\epsilon_k)$ determines a finite confinement
lifetime of QW electrons due to resonance with the continuum of
metal states. The proximity effect is represented by the second term
in the numerator of Eq.(\ref{Sigma2}). Since we assumed that
$\Gamma_{\mathbf{k}}\ll \Delta_0$ and the temperature is low, one
can neglect $\omega$ in comparison with $\Delta_0$. In this case the
gap $\Delta$ in the spectrum of QW electrons  is given by
$\Delta=\Gamma_{\mathbf{k}}(0)$ and it is  reasonable to neglect a
very weak dependence on $\mathbf{k}$, because the Fermi wave vector
is much smaller in QW  than in the metal.

The stationary effect of the electromagnetic field onto the electron
current is determined by the second-order contribution to the
conduction electron self-energy. Its 11 component in the Nambu space is
\begin{eqnarray}\label{Sigmaph}
\Sigma_{11,\alpha\beta}(\omega,\mathbf{k})&=&\sum_{v}[M_{\alpha v}(\omega_i )M^*_{\beta v}(\omega_i) G^0_{v\mathbf{k}}(\omega+\omega_i)+ \nonumber \\
&&M_{\alpha v}(-\omega_i )M^*_{\beta v}(-\omega_i) G^0_{v\mathbf{k}}(\omega-\omega_i)]\,,
\end{eqnarray}
where the unperturbed valence-band Green's functions are
$G^0_{v\mathbf{k}}(\omega)=(\omega-\epsilon_{v\mathbf{k}})^{-1}$.
From the above definition of Gor'kov-Nambu operators,  $\Sigma_{22}$
can be expressed as
$\Sigma_{22,\alpha\beta}(\omega,\mathbf{k})=-\Sigma_{11,\bar{\beta}\bar{\alpha}}(-\omega,-\mathbf{k})$.
The valence-band wave functions who determine the matrix elements in
Eq. (\ref{Sigmaph}) are  linear combinations of the functions
$|J\rangle |\sigma\rangle$, where $J=X,Y,$ or $Z$ denote $l=1$
orbitals  and $\sigma$ is the spin index. Therefore, depending on
the spin variables $\alpha$ and $\beta$ in Eq.(\ref{Sigmaph}), the
electromagnetic field enters as various combinations of
$\mathbf{A}\times\mathbf{A}^*$, $\mathbf{A}\cdot\mathbf{A}^*$ and
$A_zA_z^*$. The circular photogalvanic effect is associated with
$\mathbf{P}=i\mathbf{A}\times\mathbf{A}^*$. In general, its
contribution to  Eq. (\ref{Sigmaph}) has the form \cite{pherev}
$C\mathbf{P}\cdot \bm{\sigma}$, where the  factor $C$ contains
resonance denominators. The resonance detuning, however, was assumed
to be much larger than $\Delta$ and $h_{\mathbf{k}}$. Since $\omega
\sim$ max($\Delta,h_{\mathbf{k}},k_BT)$, a weak dependence of the
self-energy on $\omega$ can be ignored and $\Sigma$ becomes
\begin{equation}\label{Sigmaph2}
\Sigma(\omega,\mathbf{k})=\tau_3 (\mathbf{P}\cdot \bm{\sigma})C_{\mathbf{k}}.
\end{equation}

The electric current density will be calculated in the leading order
with respect to $\Sigma(\omega,\mathbf{k})$. Hence, the
corresponding correction to the Green's function of conduction
electrons in Eq. (\ref{J}) can be written as
\begin{equation}\label{expansion}
\delta G^{r(a)}_{\mathbf{k}}(\omega)=G^{0r(a)}_{\mathbf{k}}(\omega)
\Sigma(\omega,\mathbf{k})G^{0r(a)}_{\mathbf{k}}(\omega)\,,
\end{equation}
The unperturbed functions, in their turn, are given by
\begin{equation}\label{G0}
G^{0r(a)}_{\mathbf{k}}(\omega)=(\omega-\tau_3\epsilon_{c\mathbf{k}}-\bm{\sigma}\cdot \mathbf{h}_{\mathbf{k}}-\Sigma^{r(a)}_t(\omega,\mathbf{k}) \pm i\delta)^{-1}\,,
\end{equation}
where, as was mentioned above, $\Sigma^{r(a)}_t(\omega,\mathbf{k}) \simeq -\tau_1\sigma_z \Delta$.

The Green's functions of valence electrons in Eq. (\ref{J}) are
calculated in a similar way. The electromagnetic field contributes
to their self-energy through the same matrix elements as in Eq.
(\ref{Sigmaph}), with the intermediate states coming from  the
conduction band. It is easy to show that
 \begin{equation}\label{valence}
\sum_vTr\left[\mathbf{j}_{v\mathbf{k}}G_{v\mathbf{k}}(\omega)\right]=Tr\left[G^{0}_{\mathbf{k}}(\omega)\frac{\partial\Sigma(\omega,\mathbf{k})}{\partial \mathbf{k}}\right]\,.
\end{equation}

Let us choose the $x$- axis  parallel to the $xy$ projection of
$\mathbf{P}$. For simplicity, the conduction and valence bands are
assumed to be isotropic and SOI is taken in the form of the Rashba
interaction, where $h_x=\gamma k_y, h_y=-\gamma k_x$. Substituting
Eq.(\ref{valence}) and Eq. (\ref{expansion}) into Eq. (\ref{J}),
after some algebra one arrives to $J_x=0$ and
\begin{eqnarray}\label{result}
J_y&=&\frac{eN_F\gamma  C_{k_F}P_x}{2}\int d\epsilon
\left[\frac{\Delta^2}{E(\epsilon^2-h^2)} \tanh\frac{E}{k_B T}-
\nonumber  \right. \\ &&  \left. \frac{b}{2k_B T}\left(\cosh^{-2}
\frac{E}{k_B T}-\cosh^{-2}\frac{\epsilon}{k_B T}\right)\right]
\end{eqnarray}
where $E=\sqrt{\epsilon^2+\Delta^2}$, $b=(1+ d\ln C_k/d\ln
k)_{k=k_F}$ and $h=\gamma k_F$.  As expected, the equilibrium CPE
turns to 0  in the normal state \cite{Belinicher}. One can
immediately see it from  Eq. (\ref{result}) at $\Delta \rightarrow
0$. In the opposite limit of the large superconducting gap $\Delta
\gg k_B T$ the current reaches its maximum magnitude. In this case
$\cosh^{-2}(E/k_B T)$ in the integrand  is exponentially small and
in the leading approximation $J_y$ is given by
\begin{equation}\label{final}
J_y=\frac{eN_F\gamma  C_{k_F}P_x}{2}\left(\frac{\zeta^2 -1}{ \zeta}\ln \left|\frac{1-\zeta}{1+\zeta}\right| +2b\right)\,,
\end{equation}
where $\zeta=\sqrt{1+(\Delta^2/h^2)}$. We note that the expression
in brackets smoothly varies between $2b -2$, at $h \ll \Delta$ and
$2b$, at $h \gg \Delta$. Therefore,  if the light frequency is not
too close to the inter-band transition energy, one gets $b\sim 1$
and this expression is of the order of unity.

It is important to note that CPE is determined by the dependence of
the electron-photon second-order scattering on the electron spins.
The corresponding self-energy $\Sigma(\omega,\mathbf{k})$ results in
the spin orientation of electrons in the direction parallel to
$\mathbf{P}$ and, being combined with SOI, gives rise to the
electric current. A basic distinction of the equilibrium CPE in
superconductors is that at low temperatures there are no
single-particle spins to be polarized. Their role, however,  is
played by triplet Cooper pairs, which admix to singlets due to SOI.
This situation resembles the effect of the Zeeman interaction, which
together with SOI also leads to the electric current in
superconducting systems \cite{Malsh}.

Some comments are needed regarding the vector $\mathbf{P}$ in Eqs.
(\ref{Sigmaph2},\ref{result}). In the considered geometry (Fig.1),
the electromagnetic field near the QW is a sum of the incident
and reflected waves. Therefore, in the presence of highly reflecting metal, this vector is strongly modified with respect to its plane-wave value. Let the incident and reflected
light beams in  the $x^{\prime}z$ plane. Near QW the circularly
polarized incident wave has the electric field components
$E^i_{z}=-\sin\theta E_0, E^i_{x^{\prime}}=\cos\theta E_0$ and
$E^i_{y^{\prime}}=i E_0$, where $\theta$ is the incident angle  at the metal-semiconductor interface. In general, the total electric field $\mathbf{E}$ near the interface
may be expressed in terms of the surface impedance
$Z(\omega_i)$ . From these well known expressions \cite{Landau}, the
in-plane components of the vector
$\mathbf{P}=i(c^2/\omega^2_i)(\mathbf{E}\times\mathbf{E}^*)$ may be
easy calculated:
\begin{eqnarray}\label{Pxy}
P_{x^{\prime}}&=&|E_0|^2\frac{8c^2}{\omega^2_i}\sin\theta\cos^2\theta Re\left(\frac{Z}{Z^*+\cos\theta}\right)\,, \nonumber \\
P_{y^{\prime}}&=&-|E_0|^2\frac{8c^2}{\omega^2_i}\frac{\sin\theta\cos^2\theta}{|Z+\cos\theta|^2}Im Z\,,
\end{eqnarray}
where the  terms $Z\cos\theta$ have been neglected, taking into
account that the impedance of a highly reflective and thick enough
metal film is small. In the range of not too small frequencies, $Z$
is represented by its Fresnel expression
$Z(\omega_i)=\sqrt{-i\omega_i \epsilon_s(\omega_i)/4\pi
\sigma(\omega_i)}$, where $\sigma(\omega_i)$ and
$\epsilon_s(\omega_i)$ are the metal conductivity and dielectric
function of the semiconductor film, respectively. At $\omega_i \gg
\Delta_0$ the former is almost the same as  that of the normal
metal. Both, $\epsilon_s(\omega_i)$ and $\sigma(\omega_i)$ depend on
the frequency and many other factors. Therefore, it is useful to
consider typical situations. In the case of a highly reflective
metal and $\epsilon_s(\omega_i) >0$ we have $Re(Z) \ll Im(Z)$.
Hence, Eq.(\ref{Pxy})  gives the vector $\mathbf{P}$ perpendicular
to the $x^{\prime}z$ plane. Since the $x$-axis in Eq. (\ref{result})
has been chosen parallel to $\mathbf{P}$, this means that the
electric current will be directed parallel to $x^{\prime}$-axis. At
the same time, for relatively low reflective metals the real and
imaginary parts of $Z$ will be of the same order of magnitude.
Therefore, the vector $\mathbf{P}$ and, hence, the electric current
will be directed arbitrary in the $x^{\prime}y^{\prime}$ plane,
their directions varying  with the light frequency. It should be
noted that a setup different from Fig. 1  might be chosen. For
example, the light beam can be incident from the superconductor
side, penetrating through a semi-transparent film whose thickness is
less than the skin-layer depth. In this case the vector $\mathbf{P}$
will have quite different from Eq. (\ref{Pxy}) components.

For an order-of-magnitude evaluation of CPE, the factor $C$ may be
approximated by its bulk expression \cite{pherev}, that is valid far
enough from the resonance:
\begin{equation} \label{C}
C_{k}=\frac{2e^2p_{cv}^2}{3m^2c^2}\omega_1\left( \frac{1}{E_g^2-\omega_1^2}-\frac{1}{(E_g+\Delta_{so})^2-\omega_1^2} \right)\,,
\end{equation}
where $E_g$ is the semiconductor energy gap and $\Delta_{so}$
denotes the split-off energy. Taking $\omega_1 \sim E_g \sim
\Delta_{so} \sim$ 0.4 eV (InAs), we get $C \sim (e^2/m c^2)
(p_{cv}^2/m E_g) \sim (e^2/m c^2)(m/m^*)$, where $m^*$ is the
effective electron mass.  The Rashba parameter  $\gamma $ can vary
depending on QW characteristics. For InAs it can be more than
10$^{-11}$ eVm \cite{Winkler}. We take $\gamma=$10$^{-12}$ eVm,
fixing thus $h$ around 1meV at $k_F=10^6$ cm$^{-1}$, that is much
less than the Fermi energy of QW electrons. At $P_x \sim
c^2|E_i|^2/\omega^2_i$ and the moderate electric field strength
$E_i=10^4$ V/cm,  Eq.(\ref{final}) gives $J_y \sim1\mu$A/cm. This
value will increase considerably at resonance conditions, by the
factor $E_g/\delta_{res}$, where $\delta_{res}$ is a detuning from
the resonance. The above theory restricts $\delta_{res}$ by a much
larger value than  the electron and hole spin-orbit splittings. In
the safe range of relatively large detunings $\sim$10meV, the
resonance enhancement factor is about 40 for InAs based QW.

A really dramatic enhancement can be reached, if the incident light
intensity is periodically modulated with a microwave frequency that
is close to an eigenfrequency of a quantum circuit incorporating the
considered superconducting system. In principle, such a modulation
could be provided by the Rabi splitting of an optical cavity mode
strongly coupled to atomic ensembles having spin resolved electron
transitions. In this way CPE gives rise to a nonlinear interaction
of optical cavity fields + associated atomic ensembles with quantum
circuits.

In conclusion, the electric current  induced by the circular
photogalvanic effect has been calculated in the case of  a
superconducting system.  It has been shown that, unlike CPE in a
normal system,  this effect becomes possible without driving
electrons out of the thermodynamic equilibrium and without mediation
of the impurity or phonon scattering, as well as intersubband
optical transitions. This phenomenon has been considered for a
noncentrosymmetric semiconductor QW, where superconductivity is
induced by the proximity effect and the spin-orbit interaction is
represented by the Rashba term. In the considered setup the
magnitude  and direction of the electric current can be varied by
changing the incident light frequency and polarization.

I acknowledge a support from the Russian Academy of Science program
"Spin dependent phenomena in solids and spintronics".

\end{document}